\documentstyle[aps,preprint,prc]{revtex}

\begin{document}

\begin{center}
{\Large {\bf Staggering of the Nuclear Charge Radii in a Superfluid Model
with Good Particle Number}}\\

\vspace{.8cm}

{\bf Mihai Horoi$^{1,2}$\\}

\vspace{.8cm}

{\em
$^{1}$National Superconducting Cyclotron Laboratory,
East Lansing, MI 48824\\

$^{2}$Institute of Atomic Physics, Bucharest Romania\\}

\end{center}

\vspace{2cm}

\begin{abstract}
A simple superfluid model with an effective four body interaction of
monopole pairing type is used to explain the staggering of the charge
radii in the isotope chains. The contribution of deformation and of the
particle number projection are analyzed for the Sn isotopes. Good results are
obtained for the staggering parameters and neutron pairing energies.
\\

\vspace*{1cm}


\end{abstract}
\newpage

Detailed studies of the nuclear charge radii
exhibit significant deviations from an average dependence of the mass
number \cite{ANCS}. The systematic study of these deviation \cite{ANCS}
revealed two
important effects (1) the change in slope of the variation of the charge
radii versus the neutron number when crossing a magic number and (2) the
staggering of the charge radii in isotope chains (odd - even effect).
These effects are found over the whole table of nuclei and are confirmed with
increasing accuracy \cite{AN} - \cite{THOM}.

The odd-even effect exhibit some important characteristics \cite{SHE}
(a) at the beginning of a neutron shell the amplitude of the effect is
rising
and it decrease to the end of the sell;
(b) such variations in the amplitude are almost insensitive to the proton
configuration;
(c) at the end of a neutron shell an increasing of the amplitude for
lighter elements is observed.

Many attempts to explain the odd-even effect have been done during the
last 25 years but the output was often contradictory. Uher and
Sorensen \cite{UHSO} have extensively calculated the charge radii for many
isotopes chains, in the framework of pairing plus quadrupole model,
allowing for monopole and quadrupole core polarization but, they were not
able to describe the odd even staggering. Reehal and Sorensen \cite{RESO} were
able to
find agreement with the data for some cases taking into account the
influence of the neutron blocking on the ground state quadrupole
vibrations in a mixed, microscopic and liquid drop model.

Calcium isotopes have been extensively studied \cite{AN}, \cite{NOV},
\cite{TAL}, the main
mechanism involved being the core polarization, either in the Isospin
Projected Hartree Fock (IPHF) framework or using a simple parameterization
of the neutron  proton interaction in a j$^{n}$ configuration. The IPHF
mechanism \cite{COU} gives only a qualitative agreement with the data but, the
simple parameterization of Talmi \cite{TAL} nicely describe the charge radii of
calcium isotopes. However, the fitted parameters entering Talmi's formula have
not an additional physical meaning and the realistic calculation of the core
polarization contribution give a much smaller effect \cite{AN}.
 The formal extension of the j$^{n}$ formula
to the case of Pb isotopes has also received some criticism \cite{ANS}.

Some peculiar cases, connected with large deformation effects,
received particular attention. The large amplitude of the staggering in
neutron deficient Hg and Au isotopes has been attributed to an oblate -
prolate instability \cite{GIR}. For the transitional nuclei Nd, Sm and Gd,
 it is possible to obtain,
in some cases, a staggering of correct magnitude
by taking into account the zero point motion of the nuclear surface and the
effect of the h$_{11/2}$ neutron intruder state
\cite{DOBV}. Other
sophisticated calculations like the large shell model configuration for
light nuclei \cite{BRO} or the HFB model taking into account the continuum
states \cite{DOBF} for Sn isotopes, have  used to describe the
odd-even effect.

 All the above mentioned models have not been able to put in evidence an
underlying mechanism for this effect, observed with few exceptions on the
whole table of atomic nuclei. If successful, they pointed out some
specific contributions like the j$^{n}$ configuration, the prolate oblate
instability, the effect of zero point surface vibration, etc.
The decrease of the staggering toward the closure of
the neutron shell indicate that the primary mechanism which discriminate
between even and odd neutron number is pairing. This observation made by
Zawischa \cite{ZAW1} has been used to find a  mechanism that strongly
connect the neutron pairing properties with the proton properties. He
looked for an universal mechanism which  involves the
 collective properties of the neutrons and protons. The
failure of the standard nuclear models to describe the staggering indicates
that some small residual interaction can strongly contribute due to the
collectivity. For the neutron system the collective quantity must be
the pairing tensor, which have  strong odd-even variation due to the
blocking mechanism. A first order coupling of the neutron pairing tensor
to the proton field is not possible in a quasiparticle picture. The only
possibility is to introduce effective many body forces. Zawischa and
coworkers \cite{ZAW2}, \cite{ZAW3}
have introduced a separable four body interaction
between protons and neutrons within a HFB model and they were able to
correctly describe the effect for a large range of elements. A four body
interaction was indicated by the above mentioned requirements
for the underlying staggering mechanism,
and by the
importance of four body correlations to the alpha clustering
effects \cite{BUL}.
Finally, they have shown \cite{ZAW3}, \cite{ZAW4} that a three body
interaction is enough to explain the staggering amplitude.

In a recent series of  papers \cite{FC1}-\cite{FC5},
an  effective  four  body  interaction
of monopole pairing type, between pairs of protons and neutrons, have been
introduced and extensively studied. The main consequences of this new
interaction, when used in a variational approach with BCS wave functions,
are (i) the mutual induction of the superfluidity from the neutron system
to the proton one \cite{FC3}, \cite{FC5};
(ii) the appearance of local minima in the
functional energy versus the gap parameters \cite{FC2}, \cite{FC3},
which can be
interpreted as metastable (superfluid isomer) states, possible
identified \cite{FC5} with the 0$^{+}_{2}$  state  in  $^{152}$Sm.  The
property
(i) has been successfully used to described the main part of the staggering
amplitude for the Pb isotopes \cite{FC5}. This property represents, in fact,
the necessarily strong collective coupling between the superfluid neutron
system and the proton system which become superfluid. A variation in the
neutron pairing tensor, due to the blocking mechanism in the odd system,
induce a similar variation in the proton gap. The proton single particle
distribution around the Fermi level follows this variation and the
charge radii of the odd isotopes are lower than the average radii of the
neighbor even isotopes. This mechanism is in  agreement with the
qualitative explanation given by Zawischa \cite{ZAW1}. The HFB model
include much more matrix elements which can connect the long range mean
field with the pairing field; this is the reason why it includes different
type of contributions to
the staggering. For example, with three
body forces only, the main contribution comes from the coupling of the
neutron pairing tensor to the proton long range mean field \cite{ZAW3} which
lead to variations of the proton single particle wave functions, while
with four body forces the main effect is given by the staggering
of the proton pairing tensor induced by the neutron pairing tensor.

This debate was a stimulation for the present work. In a previous paper
\cite{FC5} the
Pb isotope chain has been studied because for that case
no large deformation effects are
expected due to proton magic number. The staggering
amplitude was fairly described for the N $\sim$ 126 isotopes whose
deformations are known to be small \cite{RAM}. A partial agreement was
obtained for the neutron deficient isotopes possible affected by the
deformation \cite{BEN}. In this work, the Sn isotopes are studied to further
establish the contribution of the effective four body interaction to the
staggering mechanism.
The effect of the deformation is also taken into account. The Sn isotopes
are almost spherical, but their $\beta_{2}$'s extracted from the BE2 values
\cite{RAM} are 3-4 times larger than the similar ones for Pb isotopes. However,
their variation with the neutron number is rather small and we expect no
influence of the shape fluctuation to the staggering.
It is also interesting
 to investigate the influence of the particle number violation (inherent in any
quasiparticle description) to the odd-even staggering effect. To this goal we
shall compare the results with and without the projection on good number
of particles.

In our model,
the protons and the neutrons are supposed to move in deformed single
particle orbits described by a canonical mean field H$^{mf}$, and they are
affected by a residual interaction of the usual pairing type H$^{pair}$
and a new \cite{FC1} four body interaction of monopole pairing type between
pairs of protons and neutrons H$_{4}$
\begin{equation}
H = \sum_{i=p,n} ( H^{mf}_{i} + H^{pair}_{i} ) + H_{4} \ ,
\end{equation}
where
\begin{equation}
H^{mf}_{i} = \sum_{s_{i} \sigma_{i}} E_{s_{i}} a^{+}_{s_{i} \sigma_{i}}
a_{s_{i} \sigma_{i}} \ \ ,\ \ i=p,n \ \ ,
\end{equation}
\begin{equation}
H^{pair}_{i} = -  G_{i} P^{+}_{i} P_{i} \ \ \ ,\ \ \ P_{i} = \sum_{s_{i}}
a_{s_{i}-} a_{s_{i}+} ,
\end{equation}
\begin{equation}
H_{4} =  - G_{4} P^{+}_{p} P^{+}_{n} P_{n} P_{p} \ \ ,
\end{equation}
$a^{+}_{s_{i} \sigma_{i}}$ is the creation operator of a nucleon
in a single particle state with the
sign of the angular momentum projection on the intrinsic symmetry axis
$\sigma_{i}$, $s_{i}$ representing all the other single particle quantum
numbers. The reason to preserve only this component from a full four body
interaction is that all the other components seem to give only  a
renormalization effect to the two body
strengths. This conclusion is based on the assumption that the two-body
pairing strengths are phenomenological parameters, which can incorporate terms
depending weakly on the extensive properties of the system, like $v^{4}$
(terms like $(G_{p}+G_{4p} \chi^{2}_{p})$ still have small renormalization
effects). As a consequence we keep only that part of the full four-body
interaction which has a strong dependence on the product of the proton
and neutron paring tensors.
Another reason is its simplicity that
permits to obtain  complete solutions (in a BCS approximation), which
exhibit very interesting physical properties.

The variational solutions of the proposed hamiltonian in the space of BCS
wave functions are given \cite{FC3} by the single particle densities
\begin{equation}
v^{2}_{s_{i}} (u^{2}_{s_{i}}) = \frac{1}{2} (1 - (+) \frac{E_{s_{i}} -
\lambda_{i} }{\varepsilon_{s_{i}}}) \ \ ,
\end{equation}
the quasiparticle energies
\begin{equation}
\varepsilon_{s_{i}} = [(E_{s_{i}} - \lambda_{i} )^{1/2} +
\Delta^{2}_{i}]^{1/2} \ \ ,
\end{equation}
the constraints for good average number of particles
\begin{equation}
N_{i} = \sum_{s_{i}} ( 1 - \frac{E_{s_{i}} -
\lambda_{i}}{\varepsilon_{s_{i}}})
\end{equation}
and the gap equations
\begin{equation}
(G_{p} + G_{4} \chi^{2}_{n}) \sum_{s_{p}} \frac{1}{\varepsilon_{s_{p}}} = 2
\end{equation}
\begin{equation}
(G_{n} + G_{4} \chi^{2}_{p}) \sum_{s_{n}} \frac{1}{\varepsilon_{s_{n}}} = 2
\ \ .
\end{equation}
In the above formulae the $E_{s_{i}}$ are the renormalized single particle
energies \cite{FC3}, $\lambda_{i}$ is the Fermi level for the system i,
\begin{equation}
\chi_{i} = \sum_{s_{i}} u_{s_{i}}v_{s_{i}}
\end{equation}
is the pairing tensor for a system with an even number of particles or
\begin{equation}
\chi_{i} = \sum_{s_{i} \not= so_{i}} u_{s_{i}}v_{s_{i}}
\end{equation}
for a system with an odd number of particles, when the $so_{i}$ level
is blocked. The coupling constants $G_{p}$, $G_{n}$ and $G_{4}$ can
eventually be extracted from  experimental binding energies
combination \cite{FC3} assuming that they have the following A dependence:
\begin{equation}
G_{p} = \frac{C_{p}}{A} \ \ \ ,\ \ \ G_{n}=\frac{C_{n}}{A}\ \ \ ,\ \ \
G_{4} = \frac{C_{4}}{A^{2}}\ \ .
\end{equation}
The gap equations (8) - (9) are strongly coupled by the $G_{4}$ strength and
the
collective pairing tensor of the complementary system, due to the proposed four
body interaction. This coupling has two consequences (i) the proton
system, which is in a normal phase for a magic number due to the Belyaev
condition, can become superfluid due to the contribution of the $G_{4}
\chi^{2}_{n}$ term; (ii) the staggering of the neutron pairing tensor
induced by the blocking mechanism is fairly followed by the proton gap
$\Delta_{p}$ (see Eq. (8)), the single particle densities $v^{2}_{s_{p}}$
and finally by the charge radii
\begin{equation}
<r^{2}> = \frac{2}{Z} \sum_{s_{p}} <s_{p}\vert r^{2} \vert s_{p}>
v^{2}_{s_{p}} \ \ .
\end{equation}
In particular, the charge radii are a little bit larger in an even system,
due to the fact that a larger proton gap gives larger probabilities for
protons to occupy higher single particles states with larger single
particle radii.

This simple model has the advantage that it meets the both requirements of
Zawischa and it can be extended to include quantum fluctuations.
It is interesting to try to include a 3-body effective force in this type of
dynamical calculations (not affecting the details of the single particle
wave functions but only their superfluid properties). The most simplified
form of a proton-neutron 3-body effective interaction

\[ H_{3} = -G_{pn} \sum_{s_{p} \sigma_{p}} a^{+}_{s_{p} \sigma_{p}}
a_{s_{p} \sigma_{p}} P^{+}_{n} P_{n}
-G_{np} \sum_{s_{n} \sigma_{n}} a^{+}_{s_{n} \sigma_{n}}
a_{s_{n} \sigma_{n}} P^{+}_{p} P_{p} \ , \]

\noindent
has only the effect of
 a renormalization of the pairing coupling constants
G$_{p}$(G$_{n}$) by the quantities G$_{pn}$N$_{p}$(G$_{np}$N$_{n}$)
proportional with the number of particles. This N$_{n}$ linear dependence of
G$_{p}$ cannot explain any staggering.
This represents a clear indication that the staggering induced by
3 body forces within the HFB model \cite{ZAW4} originates mainly from the
modification of the shape of the single particle wave functions
due to the nonlinear coupling of the neutron pairing tensor to the long
range mean field.

One of
the drawbacks of the BCS  wave functions consists in their
nonconservation of the particle number. Fluctuations of the particle
number give large contributions to quantities like the total energy but,
they can be important also for such delicate observables like the small
variation of the charge radii. In order to study such an effect, particle
number projected BCS wave functions are used
\begin{equation}
\vert \Psi > = P_{A} \vert BCS> = \frac{1}{2 \pi} \int_{0}^{2 \pi} d \phi
e^{i \phi(\hat{N} - A)} \prod_{s_{i}}(u_{s_{i}} + v_{s_{i}} a^{+}_{s_{i}+}
a^{+}_{s_{i}-}) \vert - > \ \ ,
\end{equation}
where $\hat{N}$ is the particle number operator. The relevant matrix elements
can be obtained
\begin{equation}
<\Psi \vert \Psi > =R^{0}_{0}\ ,
\end{equation}
\begin{equation}
<\Psi \vert a^{+}_{s}a_{s} \vert \Psi> = R^{1}_{1} (s)\ ,
\end{equation}
where $R^{m}_{n}$ functions can be analytically written, but a
recursive relation \cite{DMP}
\begin{equation}
R^{m}_{n}(s_{1}...s_{m}) = u^{2}_{s}R^{m+1}_{n}(s_{1}...s_{m}s) +
v^{2}_{s} R^{m+1}_{n+1}(s_{1}...s_{m}s)
\end{equation}
is much more suitable for computer calculations. The charge radius is
given by
\begin{equation}
<r^{2}> = \frac{2}{Z R^{0}_{0}} \sum_{s_{p}} <s_{p} \vert r^{2} \vert
s_{p}> v^{2}_{s_{p}} R^{1}_{1} (s_{p}) \ \ .
\end{equation}
The qualitative effect of the particle number projection on the BCS wave
functions is that the single particle distribution around Fermi surface is
squeezed.

Due to the fact that Sn has a proton magic number and in order to study
the effects of different parameters, it is interesting to
investigate this chain of isotopes in a spherical model. For the
calculation of the single particle wave functions and energies, a program
developed by Hird \cite{HI} has been used. It has the advantage that the
eigenvectors are obtained in terms of the tridimensional harmonic
oscillator wave function
basis for which the matrix elements are analytically known.

The single particle potential is of the Woods Saxon type, allowing
$\beta_{2}$ and $\beta_{4}$ deformations. The potential parameters are
taken from Ref. \cite{SOL}, page 21 (quoted as Cheprunov in Table 1 of Ref.
\cite{CW}).
For computational reasons (to get double degenerate levels),
the calculations have been performed with a very small deformation
\[ \beta_{2} = 0.005 \ \ \ \ ,\ \ \ \ \beta_{4} = 0.0 \ \ . \]
The results are summarized in Table 1. The staggering is
qualitatively obtained but the absolute results deviate from the
experimental values. The main difficulty comes from the fact that the
slope of variation of the mean square radii versus the mass number A is
not correctly reproduced. This difficulty has its roots in the
shell model results (pairing residual interaction neglected) as can be
extracted from the third column of Table 1. Possible explanations can be: no
enough harmonic oscillator shells included, an inaccurate A and Z dependence
of the Woods Saxon parameters, deformation effects not taken into account.
The contribution from the higher harmonic oscillator shells has been carefully
checked and found negligeable.
The parametrizations Blomqv - Wahlb, Rost and
"universal" from
Table 1 in Ref. \cite{CW} have been checked but no qualitative
improvements have been obtained. The effect of the deformation will be
discussed in the next section.

The
odd-even effect is obtained but the magnitude of the variation and the
amplitude of the
staggering are half the experimental one (see Table 1).
 In order to obtain this effect a
fixed set of two and four body pairing strengths has been used
\[ C_{p}=33.5\ MeV\ \ ,\ \ C_{n}=20.\ MeV\ \ ,\ \ C_{4}=2.\ MeV\ \ . \]
These parameters can be extracted for every nucleus
from the correlation energies
$P_{Z}$, $P_{N}$ and $P_{4}$ (defined in Ref. \cite{FC3})
with a relatively complicated procedure
described in Ref. \cite{FC3}. In this work
we use the same strengths for the whole chain of isotopes and compare their
values with the range given by the procedure described in Ref. \cite{FC3}.
The small slope of the radii versus the mass number A, force a relatively
large proton pairing strength, $C_{p}$ and a small $C_{4}$ quantity (in
Ref. \cite{FC3} the experimentally extracted $C_{4}$ in the rare earth
region are 10 times larger).

The staggering parameter \cite{UHSO}
\begin{equation}
\gamma_{A\ even}=
\frac{2[<r^{2}>_{A+1}-<r^{2}>_{A}]}{<r^{2}>_{A+2}-<r^{2}>_{A}} \ \ ,
\label{eq:stp}
\end{equation}
represents an intrinsic measure of the effect. The calculated $\gamma
_{A}$ (see Table 1) is in very good accord with the experiment, indicating
that the proposed mechanism is adequate to describe the staggering.

One can try to improve the above results by including the deformation
in the model. The static deformations for Sn isotopes are unknown but the
$\beta_{2}$ values extracted from the B(E2) values can give a fairly good
approximation if the deformation is not too small.
The $\beta_{2}$ values for the even
isotopes can be extracted from the measured B(E2) values according to the
following formula \cite{STAC}:
\begin{equation}
\beta_{2} \equiv \sqrt{<\beta^{2}_{2}>} = \sqrt{B(E2)}
\frac{4\pi}{3 Z R^{2}_{0}}\ \ ,
\end{equation}
where $R_{0}$ is usually taken as 1.4 $A^{1/3}$ fm. The extracted
$\beta_{2}$ values for the Sn isotopes are around 0.1 \cite{RAM}, 3-4 times
larger than the similar values for the Pb isotopes, and their
total variation is around 0.04 (see Table 2). In order to have an idea of
the effect of this variation of deformation on the change in the mean
square radii one can use the model of a uniformly charged deformed
nucleus \cite{SHE}
\begin{equation}
\delta <r^{2}> = \delta <r^{2}>_{sph} + <r^{2}>_{sph} \frac{5}{4\pi}
\delta <
\beta ^{2} >\ \ .
\end{equation}
According to this formula a deviation of 0.01 in $\beta _{2}$ can give,
for nuclei with A $\sim$ 120, a deviation of 0.01 in $\delta <r^{2}>$. Such
a magnitude is just the unit in the staggering. This classical result was
checked also in a microscopic calculation.

Another effect of the deformation is that it gives a better slope of
variation of the mean square radii versus the mass number A. The results
are presented in Table 2 and Fig. 1. The deformations used in the
calculations for the even nuclei have been taken from Ref.
\cite{RAM}. For the odd
nuclei the interpolated values have been used. Figure 1 indicates that the
discrepancy in slope has been reduced but
the experimental results are not completely
reproduced. The
remaining difference can come from a better A dependence of the $R_{0}$
Woods Saxon parameter.

The two and four body pairing strengths used in the calculation are
\[ C_{p} = 30.\ MeV\ \ \ ,\ \ \ C_{n} = 18.\ MeV\ \ \ ,\ \ \ C_{4} = 10.\
MeV\ \ . \]
The proton pairing strength is not yet decreased very much but, the four
body strength is near the right magnitude \cite{FC3}, \cite{FC5}.
 The magnitude of the
variation of the of the radii and the staggering amplitude are now
closer to the experimental ones (see Fig. 2). The staggering
parameter is smaller than the experimental one (see Fig. 3), indicating
that the effect is underestimated.

The effect of particle number projection is shown in Fig. 3 where the
staggering parameters with and without projection are compared. The
results consist in an increase of $\gamma_{A}$ for the projected case in
the neutron rich region. This behavior is due to the fact that the
amplitude of the staggering comes from the amplitude of variation in the
proton gap; for the neutron rich isotopes the gap is already small in the
BCS approximation but, the  projection shrinks further the proton density
around the Fermi level, making the staggering amplitude smaller.

It is interesting to see if the quoted pairing constants are able to
reproduce the experimental pairing energies,
\begin{equation}
P_{Z} = \frac{1}{2} [2{\cal E} (Z-1,N)-{\cal E} (Z,N)-{\cal E} (Z-2,N)]
\label{eq:pz}
\end{equation}
\begin{equation}
P_{N} = \frac{1}{2} [2{\cal E} (Z,N-1)-{\cal E} (Z,N)-{\cal E} (Z,N-2)]\ \ ,
\label{eq:pn}
\end{equation}
where the -$\cal{E}$ = B is the binding energy. These quantities are
compared to the experiment in Fig. 4. The neutron pairing energy is
fairly reproduced. The shell effect around A = 116, connected with a shell
closure for N = 64 is also obtained.
The experimental values for the proton pairing energies, $P_{Z}$, are larger
than 1 MeV. This give some indication that the proton system is superfluid
to some extent; this effect can be explained by the 4-body neutron-proton
coupling.
The theoretical description of the proton pairing energies is not very
accurate,
 especially in the neutron deficient region.
This could rise some doubts about the  agreement obtained for the
amplitude of the staggering. The origin of this discrepancy is due to a larger
$C_{p}$ values necessary to assure proton superfluid properties to all
even isotopes in the chain. However, if one wants to describe only the
neutron deficient part of the isotope chain with a fixed set of $C_{i}$
strengths,
one could do this with a lower value of $C_{p}$ and, as a consequence,
with a better description of the proton pairing energies.
A more accurate description of the neutron number dependence for the slope
of the radii and for the $G_{p}$, $G_{n}$, $G_{4}$ strengths (see Eq. (12))
could solve this discrepancy.

 Finally, it
is interesting to compare the absolute rms radii with the values
experimentally extracted \cite{PIL}. The results are given in the last
column of Table 2. They are in good agreement with the experiment
for the neutron deficient isotopes (compare to the data in the prelast
column of the table). Similar discrepancy have been obtained in the more
sophisticated HF calculations for Te isotopes (see Table XIV  in Ref.
\cite{SHE}).

In conclusion,
the staggering of the nuclear charge radii of the Sn isotopes has been
investigated in a simple superfluid model with an effective four body
interaction of monopole
pairing type included. The intrinsic  effect  has  been  obtained  in  a
simple quasispherical
approximation but, the absolute results deviate from the experimental values
due to the weaker dependence
of the slope of the charge radii versus the neutron number.
The contribution of the particle number projection and of the
deformation have been studied, their effects leading to better
absolute results. This simple model has two interesting features:
(i) it fulfill the qualitative physical requirements of the staggering
mechanism discussed by Zawischa; (ii) it can be extended to include
quantum fluctuations like particle number projection, RPA pairing and
quadrupole vibrations in the ground state or fluctuations in the
gauge space connected with the gaps. Its parameters have known
physical meaning and its ability to describe the staggering data, make it
a good candidate model to check the contributions of different
mechanisms to the odd-even staggering of the charge radii.

\vspace{1cm}
{\bf{Acknowledgments}}

The author would like to acknowledge support from the
NSF grant 94-03666.

\newpage

{\bf{Table Captions}}
\vspace*{1cm}
\newcounter{tab}
\begin{list}{\bf Table \arabic{tab} }{\usecounter{tab}}

\item
Relative and the variation of the
mean squared charge radii (in $fm^{2}$) for Sn isotopes. Experimantal values
are taken from Ref. \cite{HEI}.
Theoretical values obtained in the quasispherical model. Last two columns
present results for the staggering parameter, Eq. (\ref{eq:stp}).

\vspace*{6mm}

\item
Charge radii results with deformations included. First three columns
give the relative mean square charge radii in $fm^{2}$. $\beta _{2}$
values are taken from Ref. \cite{RAM}. Last two columns show the
absolute charge
radii in $fm$; experimental values from Ref. \cite{PIL}.

\end{list}

\newpage
{\bf{Figure Captions}}
\newcounter{fig}
\begin{list}{\bf Figure \arabic{fig} }{\usecounter{fig}}

\item
Relative mean squared charge radii for Sn isotopes. Experimental values
(squares and solid line)
taken from Ref. \cite{HEI}. Theoretical values (circles and solid lines)
are calculated
with four body forces and deformation included. Theoretical values calculated
with projected wave functions are denoted by pluses and
are connected by dashed lines.  All lines are
drawn to guide the eye.

\vspace*{6mm}

\item
Staggering of the mean squared charge radii. The significance of  signs and
lines is the same as in the caption to Fig. 1.

\vspace*{6mm}

\item
Theoretical staggering parameter compared with the experimental one.
The significance of  signs and
lines is the same as in the caption to Fig. 1.

\vspace*{6mm}

\item
Neutrons (left) and protons (right)
pairing energies  given by Eqs.
(\ref{eq:pn}) and (\ref{eq:pz}). Experimental values
marked by squares and solid line.
Theoretical values (circles and dashed lines)
are calculated
with four body forces and deformation included. All lines are
drawn to guide the eye.

\end{list}

\newpage

Table 1
\vspace*{3cm}
\begin{tabular}{|l|c|c|c|c|c|c|}
\hline
\multicolumn{1}{|c}{A} &
\multicolumn{1}{|c}{Exp.} &
\multicolumn{1}{|c}{Shell model} &
\multicolumn{1}{|c}{4N model} &
\multicolumn{1}{|c}{$\delta <r^{2}>_{A,A-1}$} &
\multicolumn{1}{|c}{$(\gamma_{A})_{exp}$} &
\multicolumn{1}{|c|}{$(\gamma_{A})_{4N}$} \\
\hline \hline
110 & -0.638 & -0.439 & -0.360 &   & 0.736 & 0.876\\
111 & -0.586 & -0.388 & -0.323 & 0.037 & &\\
112 & -0.497 & -0.339 & -0.275 & 0.048 & 0.908 & 0.806\\
113 & -0.438 & -0.293 & -0.244 & 0.031 & & \\
114 & -0.367 & -0.246 & -0.198 & 0.046 & 0.687 & 0.706\\
115 & -0.322 & -0.202 & -0.173 & 0.025 & & \\
116 & -0.236 & -0.159 & -0.127 & 0.046 & 0.758 & 0.718\\
117 & -0.189 & -0.118 & -0.103 & 0.024 & & \\
118 & -0.112 & -0.077 & -0.060 & 0.043 & 0.750 & 0.707\\
119 & -0.070 & -0.038 & -0.039 & 0.021 & & \\
120 &  0.0   &  0.0   &  0.0   & 0.039 & 0.832 & 0.732\\
121 &  0.042 &  0.038 &  0.019 & 0.019 & & \\
122 &  0.101 &  0.074 &  0.052 & 0.033 & 0.725 & 0.723\\
123 &  0.134 &  0.109 &  0.069 & 0.017 & &\\
124 &  0.192 &  0.144 &  0.099 & 0.030 & &\\
125 &  0.225 &  0.178 &  0.112 & 0.013 & &\\
\hline
\end{tabular}

\newpage

Table 2
\vspace*{3cm}
\begin{tabular}{|l|c|c|c|c|c|c|}
\hline
\multicolumn{1}{|c}{A} &
\multicolumn{1}{|c}{Exp.} &
\multicolumn{1}{|c}{4N model} &
\multicolumn{1}{|c}{Projected} &
\multicolumn{1}{|c}{$\beta _{2}$} &
\multicolumn{1}{|c}{rms exp.} &
\multicolumn{1}{|c|}{rms 4N model} \\
\hline \hline
110 & -0.638 & -0.504 & -0.512 & 0.126  &        & 4.5820 \\
111 & -0.586 & -0.462 & -0.467 & 0.1245 &        & 4.5866 \\
112 & -0.497 & -0.386 & -0.390 & 0.1227 & 4.5958 & 4.5949 \\
113 & -0.438 & -0.329 & -0.328 & 0.1208 &        & 4.6011 \\
114 & -0.367 & -0.257 & -0.259 & 0.119  & 4.6103 & 4.6089 \\
115 & -0.322 & -0.230 & -0.229 & 0.1155 &        & 4.6118 \\
116 & -0.236 & -0.151 & -0.159 & 0.1118 & 4.6261 & 4.6203 \\
117 & -0.189 & -0.135 & -0.135 & 0.111  & 4.6320 & 4.6220 \\
118 & -0.112 & -0.074 & -0.077 & 0.1106 & 4.6395 & 4.6287 \\
119 & -0.070 & -0.057 & -0.054 & 0.109  & 4.6448 & 4.6306 \\
120 &  0.0   &  0.0   &  0.0   & 0.1075 & 4.6522 & 4.6367 \\
121 &  0.042 &  0.015 &  0.024 & 0.1055 &        & 4.6383 \\
122 &  0.101 &  0.070 &  0.072 & 0.1036 & 4.6633 & 4.6442 \\
123 &  0.134 &  0.086 &  0.106 & 0.100  &        & 4.6460 \\
124 &  0.192 &  0.135 &  0.150 & 0.0953 & 4.6736 & 4.6512 \\
125 &  0.225 &  0.165 &  0.191 & 0.092  &        & 4.6545 \\
\hline
\end{tabular}

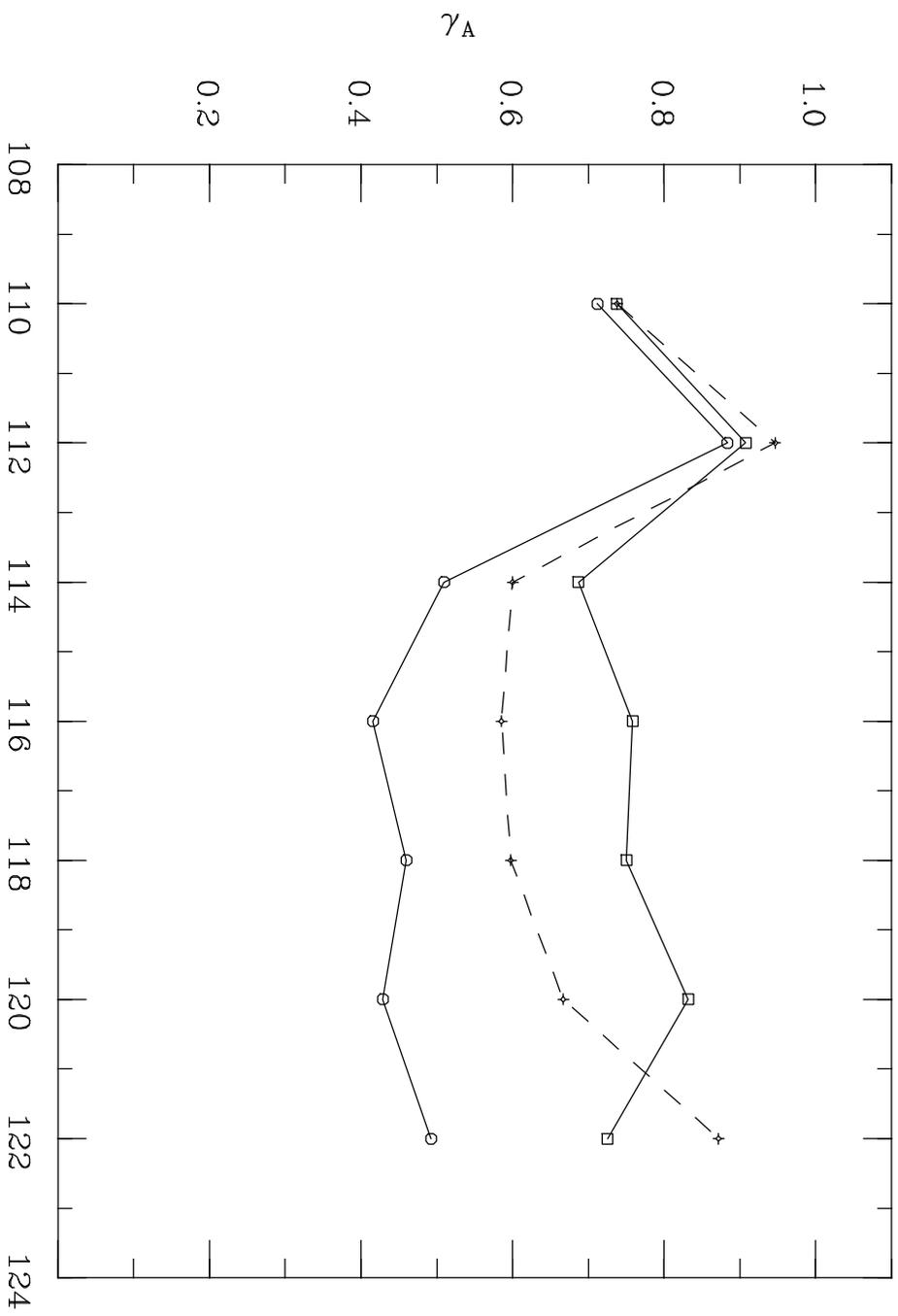
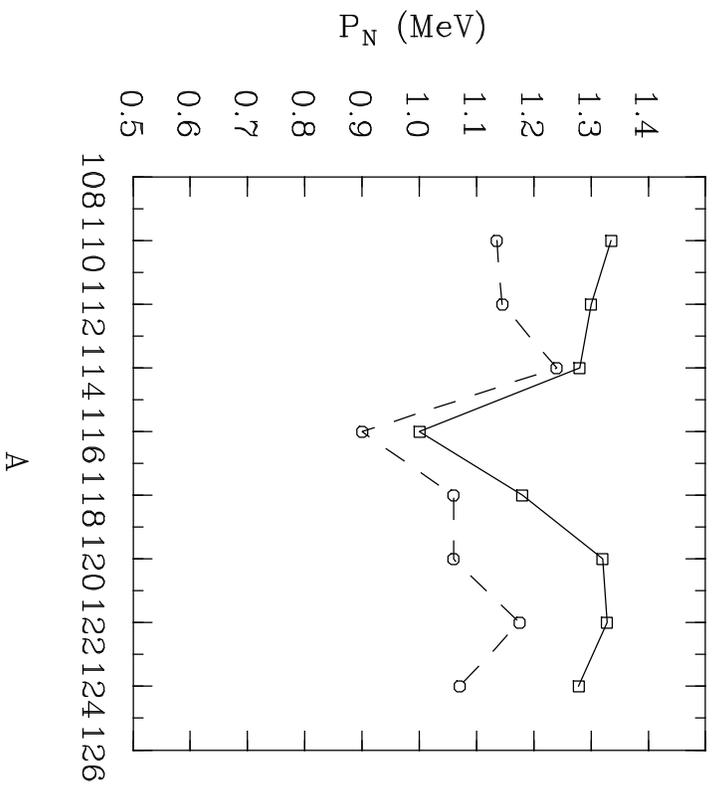
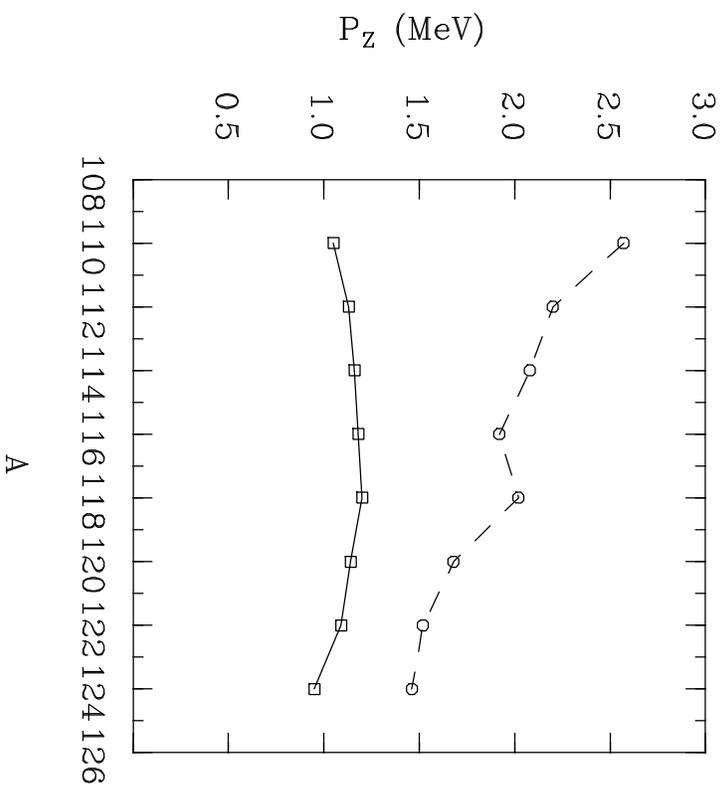
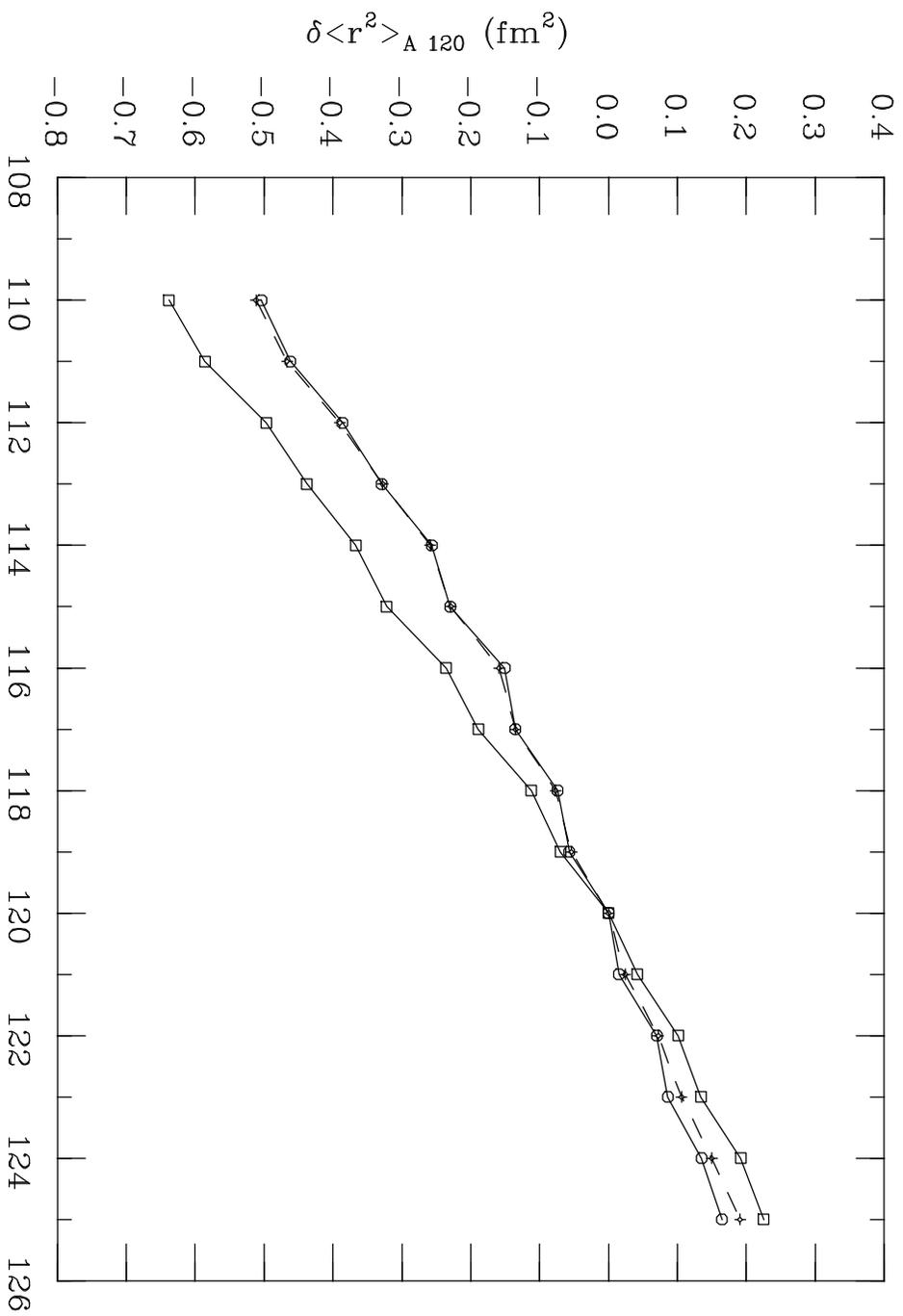
\end{document}